\documentclass[3p,times,procedia]{elsarticle}
\usepackage{nupha_ecrc}
\volume{00}

\firstpage{1}

\journalname{Nuclear Physics A}
\runauth{}

\jid{nupha}

\jnltitlelogo{Nuclear Physics A}

\usepackage{amssymb}
\usepackage[figuresright]{rotating}
\usepackage[utf8]{inputenc}

 \newcommand{\Jpsi} {J/$\psi$}

\newcommand{\ee}{$\mathrm{e}^+\mathrm{e}^-$}

 \newcommand{\Ups}{$\Upsilon$}

 \newcommand{\Tchem} {\ensuremath{T_{\rm chem}}}
 \newcommand {\PT} {\ensuremath{p_T}}

 \newcommand{\RAA} {\ensuremath{R_{AA}}}

\newcommand{\he}{$^{3}{\mathrm{He}}$}
\newcommand{\hyp}    {$^{3}_{\Lambda}\mathrm H$}

\begin{document}

\begin{frontmatter}
\dochead{XXVIth International Conference on Ultrarelativistic Nucleus-Nucleus Collisions\\ (Quark Matter 2017)}

\title{QM2017: Status and Key open Questions in Ultra-Relativistic Heavy-Ion Physics}
',

\author{Jurgen Schukraft}

\address{CERN, Div. EP, CH-1211 Geneva 23, Switzerland;   schukraft@cern.ch}

\begin{abstract}
Almost exactly 3 decades ago, in the fall of 1986, the era of experimental ultra-relativistic (\emph{E/m $\gg 1$}) heavy ion physics started simultaneously at the SPS at CERN and the AGS at Brookhaven with first beams of light Oxygen ions at fixed target energies of 200 GeV/A and 14.6 GeV/A, respectively. The event was announced by CERN \cite{cernpress1,cernpress2} with the usual superlatives ''Break new ground.., World Record Energy ..", but also with the information that "up to 400 particles were created per collision" and that "over 300 physicists .. analyzing the data .. [try] to find out whether the famous quark-gluon plasma really has been achieved". One would have thought that with almost one physicist per particle, this would have been figured  out rather quickly. However, as we know today, 30 years and 21 Quark Matter conferences later, the study of dense and hot matter, of the strong interaction in the non-perturbative regime, has been a long and winding road. The journey was much more difficult and time consuming, but also much more interesting and rewarding, than anyone could have anticipated, with many twists, some dead ends, and a never-ending string of surprises. This $30^{th}$ anniversary of heavy ion physics, and the start of the 26$^{th}$ Quark Matter in Chicago, is a good opportunity to look back and mention a few of the major results from each of the three eras (fixed target/RHIC/LHC), along with some of the answers they have provided us and some of the key questions which remain to be solved.
\end{abstract}



\end{frontmatter}


\section{3x3 Key Results from 30 years of Heavy Ion Physics}

In early 2000, shortly before RHIC start-up, CERN summarised the main results of its SPS fixed target program in a notable  announcement  \emph{"New State of Matter created at CERN"} which stated that \emph{" ..compelling evidence for a new state of matter.."} had been found which \emph{"..features many of the characteristics of the theoretically predicted quark-gluon plasma.."} \cite{press2000}. Amongst the major findings was the \textbf{strangeness enhancement}, as predicted for thermal production in equilibrated matter/QGP, the \textbf{anomalous \Jpsi~ suppression}, expected as a deconfinement signal, and the enhanced production of \textbf{low mass lepton pairs} around the $\rho$ mass region, potentially indicating the onset of chiral symmetry restoration\footnote{
Well known references will not be listed explicitly; they can be found in recent heavy ion reviews, e.g. \cite{Schukraft:2015dna, Muller:2006ee, Braun-Munzinger:2015hba, Muller:2012zq, Roland:2014jsa, Armesto:2015ioy}.}.
While this account was not unanimously accepted at the time (nor is it today \cite{ Tannenbaum:2014tea}), with hindsight, the results have all stood the test of time; in later years they were confirmed and improved upon both at the SPS (in particular the low mass lepton pair measurements \cite{Arnaldi:2008fw}) and at the RHIC beam energy scan \cite{ Luo:2015doi}. However, it was not really understood at the time that the "new state of matter" was anything like the predicted quark-gluon plasma, widely presumed to be a quasi-free gas of weakly interacting partons. Following longstanding European tradition, a discovery indeed had been made, but wrongly announced to be the "Islands of India beyond the Ganges".

Amongst the many striking results from RHIC was the ever increasing \textbf{elliptic flow}, reaching what was then (wrongly) thought to be the maximum value possible for an ideal liquid with vanishing shear viscosity; the suppression of high \PT~particles, caused by energy loss or \textbf{"jet quenching"} in the hot and dense matter, 
and a charge dependent two-particle correlation called \textbf{Chiral Magnetic Effect} (CME) potentially related to the QCD chiral anomaly. The first two results established that the "New World" was actually a \emph{".. state of hot, dense matter .. quite different and even more remarkable than had been predicted .."} \cite{bnlpress}, and thus gave rise to today's standard model of heavy ion physics: the sQGP as a strongly interacting (almost) perfect liquid. The status of the CME signal remains unsettled as of today; it may yet transform into a most exciting "beyond the standard model" revelation of local strong parity violation, or be exposed as a tricky background mix of elliptic flow and local charge conservation.

The strength of LHC was thought to be primarily in precision measurements of hard probes\cite{Schukraft:2006nt}, aided by its high luminosity and high energy. After a furious start in 2010 \cite{cernpress2010}, which quickly rediscovered and confirmed some of the major RHIC results, it certainly delivered in this respect, for example with pioneering, and then increasingly detailed and differential, measurements of \textbf{jet modifications} and heavy-flavour matter interactions and the resulting transport coefficients. However, it also had its share of discoveries, for example the solution to the \Jpsi~puzzle, i.e. \textbf{\Jpsi~regeneration}, which can explain the energy independence of the measured \Jpsi~suppression going from SPS to RHIC. And then there was the totally unexpected: the surprise discovery of a \textbf{"Near Side Ridge"} correlation in small systems (pp/pA); possibly a manifestation of subtle quantum effects from an initial state Colour Glass Condensate (CGC), or, more parsimonious, the smooth continuation of heavy ion phenomena to small systems and low density. 

The following chapters will come back in more detail to each of these 9 key results - 3 from each era - and briefly mentioning some of the lessons learned and some of the questions left unanswered or newly revealed. This article, based on the opening talk given at the 2017 Quark Matter Conference in Chicago, is not meant to be a learned and comprehensive review article for non-experts (it isn't), but rather a personal, totally biased, but hopefully stimulating and certainly provocative collection of half-digested ideas and questions targeted at the heavy ion community. 
Some of them may find an immediate answer at this conference, others are probably ill informed or ill posed, while the remainder may be profound or at least difficult enough to keep us busy for a number of years to come. 

\section{Key Questions}
\subsection{Statistical/Thermal Particle Production}
By far the most economic and successful model of particle production (hadronisation)  in high energy collisions is the thermal or statistical model, which describes dozens of particle ratios in different high energy reactions (from \ee~over pp/pA to AA) with very few free parameters;  in fact a single one - the chemical freeze-out temperature \Tchem~ - in nuclear collisions at LHC. The underlying idea of statistical equilibrium distributions is simple and intuitive, but despite decades of success and a number of plausible ideas, there is   not a single experimental result which would tell us conclusively \emph{how} or \emph{why} it works as well as it does. The two most often cited underlying mechanisms can be summarized as "evolving into equilibrium", i.e. reaching equilibrium in a finite time via reactions and detailed balance in either the hadronic or the partonic phase, and "born into equilibrium", which stipulates that averaged over many possible reaction channels transition amplitudes are dominated by the phase space factor. In either case, the resulting distribution of particle species fills the available phase space uniformly. However, having reached this state of maximal entropy and minimal information, dynamics has been replaced by thermo-dynamics and the system lost any trace of when or how it arrived at this point. This is very unfortunate, because understanding the dynamics of hadronisation is important both theoretically, as it links the confined hadronic and the deconfined partonic phases, and for any number of practical reasons (viscous corrections at freeze-out, quarkonia regeneration, constituent quark scaling, soft-hard recombination, ..). If equilibrium is the problem, hiding dynamics, looking for out-of-equilibrium remnants may be the solution to uncover dynamics. In an evolving system, whether on its way into or out of equilibrium, relaxation times and reaction rates for different processes are directly related to the corresponding cross sections. Therefore, given the large variety of reaction channels and hadronic cross sections, it is not a question of \emph{if} different particles will freeze-out at different times and temperatures, but only a question of \emph{how big} the effect is in each individual case. Identifying such a freeze-out hierarchy in certain particle ratios could make, not break, the case for hadro-chemistry!

One way of looking for "dynamics at work" are the recent studies of size and density (dN/dy) dependence of particle ratios in pp and pA, some of them to be shown at this conference. Without any initial or final state interactions (i.e. additional dynamics), there seems no obvious mechanism how the "born into phase space" idea can describe the significant change of particles ratios from MinBias pp to central AA.
The observed pattern is intriguing and non-trivial also in the context of the statistical model, as the strangeness canonical version (exact Event-by-Event (EbE) strangeness conservation) fails the crucial $\phi/\pi$ test \cite{alex}. In addition, at LHC ($\mu_B$=0), baryon conservation should be as relevant as strangeness conservation. However, no suppression is seen for protons while one could naively expect the effect to be roughly similar to the one measured for kaons or lambdas (somebody should do the actual calculation!). A phenomenological approach of combining the statistical model with a Core-Corona recipe to describe the dN/dy dependence also achieves only a rather modest accuracy, despite introducing quite a few additional free parameters\cite{alex}.

Another hint of non-equilibrium may be the surprisingly low p/$\pi$-ratio at LHC; despite pulling down by almost 10 MeV the chemical freeze-out temperature compared to pre-LHC fits - which leads to a slight tension with some Hyperon abundances which seem to prefer the original \Tchem~ - the  p/$\pi$-ratio remains well below most predictions. Amongst a number of possible explanations, baryon final state annihilation (i.e. late baryon freeze-out) is a leading but not the only contender \cite{Floris:2014pta}. The well-established density dependence of some resonance ratios ($\rm K^*, \Lambda^*$) is presumably of a different nature - decay followed by elastic scattering of the decay products. Resonance ratios are therefore a sensitive measure of the final state interaction taking place \emph{after} hadronisation; they are well described by transport codes (EPOS, URQMD) and consequently strongly disfavour models with a single (\Tchem =$T_{\rm kin}$) chemical and kinetic freeze-out. 

A puzzle, around since more than 30 years, has  returned with a vengeance at the colliders: the production of light (anti)nuclei (d, t,  $^{3}\rm{He}$,\hyp, $^{4}\rm{He}$). Two models compete to explain their yields and momentum distributions: 
Thermal/statistical production, like any other hadron, right at the phase boundary followed by kinetic expansion, or coalescence of nucleons into nuclei at around kinetic freeze-out. While it is often claimed that both models give similar yields, this is actually only correct as an order of magnitude estimate: typical ab initio calculations (as compared to phenomenological fits with momentum and/or space cut-offs parameters) differ usually by factors between two and ten \cite{Cho:2017dcy, Mrowczynski:2016xqm}. This is actually to be expected because the parameters governing the yield in both cases are very different and have no obvious relation to each other: Nuclear masses and \Tchem~in one model, phase space distributions of constituents and nuclear wave functions (i.e. size) in the other.  A case in point is the ratio between \he~ and \hyp : This depends in the thermal model essentially on temperature and mass difference, whereas in coalescence calculations the very different binding energy, and therefore different size of the two nuclei ($\approx$ 2 fm vs $\approx$ 5 fm), leads to an additional factor of five suppression for the larger hypertriton \cite{Sun:2015ulc}, with no equivalence in the thermal model. Therefore, even if it might be possible with some tweaking to get agreement between the two models at the level of the current experimental precision, of order 20\% for the most abundant nuclei, this would seem to be at best accidental and more likely spurious.

Both yields and phase space distribution of light nuclei at LHC agree with amazing precision with the thermal + kinetic model (with all parameters fixed via other hadrons). The puzzle however is how these lightly bound nuclei survive the kinetic scattering phase intact until kinetic freeze-out, given the small energy threshold (typical binding energy is MeV or less) and the huge cross sections (the inelastic cross section for pion deuteron break-up exceeds 100mb \cite{ PhysRevLett.48.577} and is actually larger than the elastic cross section!).   Note that neither entropy conservation nor detailed balance can serve as explanation: The former only restates the problem in that \emph{if} the expansion is isentropic (in the baryon sector), \emph{then} the nuclei/nucleon ratio does not change; it does not explain why entropy does not increase, as it should in the presence of significant inelastic processes. And detailed balance (i.e. the back reaction) can only drive yields towards the local thermal equilibrium, i.e. in an expanding and cooling medium the nuclei abundance would have to drop in line with the decreasing temperature. For the time being, nucleosynthesis remains as puzzling as ever, but detailed transport calculations (including inelastic reactions with realistic break-up cross sections) are urgently needed to make the discussion quantitative.

It is too early to tell if we really see in the data an approach to (or a falling out of) equilibrium or some other hints of the actual dynamics behind the unreasonable success of the statistical model. However, the recent precision measurements of particle ratios in both small and large systems have definitely revived interest in a topic which was thought by some to be largely understood and essentially exhausted and 
led to a number of fresh ideas concerning hadronisation and particle production
(e.g. Colour Ropes \cite{Bierlich:2014xba}, 
thermal string decay \cite{Fischer:2016zzs}, 
Unruh radiation \cite{Castorina:2007eb}, 
string fusion \cite{ Amelin:2001sk}, 
flavor differential FO \cite{Bellwied:2013cta}, 
missing resonances \cite{Alba:2017cbr},..). And they also make a strong case to re-measure these ratios with state-of-the art precision at RHIC to look for tell-tale changes with energy/particle density.
 
\subsection{Thermal Radiation and Chiral Symmetry}
While some hints were found at the SPS of excess direct photons roughly consistent with expectations from thermal radiation, the first significant observation was made at RHIC. The inverse slopes of the direct photon excess ($\approx 200$ MeV at RHIC and $\approx 300$ MeV at LHC) can be related to a temperature only in a model dependent way (because of space-time averaging and flow blue shift), but indicate initial temperatures of order T$_0 \approx$ 300 - 500 MeV, far above T$_c$. The most precise measurement (slope $200 \pm 10$ MeV) at this point still comes from the muon pair invariant mass spectrum at the SPS, which is, by definition, Lorentz invariant and therefore not blue-shifted. A number of key issues remain to be addressed for direct photons: on the experimental side primarily reducing errors (stat+syst) to confirm that $T_0$ indeed increases with energy between RHIC and LHC, as the current evidence is only marginally significant; on  the theory side more detailed calculations including e.g. medium effects on rates and state-of-the-art space-time evolution. Describing the direct photon $v_2$, which is within (sizeable!) errors as large as the $v_2$ of charged particles, remains a major challenge; even unorthodox choices of model parameters (eg for the transition temperature T$_c$) often lead to only a moderate agreement between data and theory.

The situation in the low mass lepton pair region around the $\rho$ meson is more satisfactory, in particular since the two RHIC experiments now have consistent results. A coherent theory describes the high precision data at the SPS as well as the RHIC results. The connection between the observed broadening of the $\rho$ in hot hadronic matter and chiral symmetry restoration in the QGP is however still more of a conjecture than a solid theoretical derivation. On the experimental side, precision results from LHC are still missing, and may have to wait a few years for detector upgrades. Likewise needed is better data on the intermediate mass region (thermal radiation, charm) from both RHIC and LHC, and ideally even in pp/pA to see if the collective features observed in small systems go hand in hand with "hot matter". On the long run, new fixed target experiments (CBM, NA60+), which benefit from high luminosity and clean muon identification, should deliver quality data in the high baryon density regime. 

\subsection{Quarkonium Suppression and Deconfinement}
The history of Quarkonia suppression as a "well calibrated smoking gun" for deconfinement can best be summarized as long and tortured: Predicted in 1986 shortly before first collisions, \Jpsi~suppression was discovered in the data right away, only to be un-discovered a few years later as a cold nuclear matter effect, just in time to be re-discovered as "anomalous" with the Pb beams, and to then linger in limbo as the SPS $\approx$ RHIC puzzle (i.e. the fact that \Jpsi~suppression was similar at both machines despite the significant difference in energy). Redemption, so it is hoped, may have finally arrived in the form of an increased \Jpsi~\RAA~at LHC, whose \PT~and centrality dependence neatly shows the predicted hallmarks of \Jpsi~regeneration during hadronisation. Ironically, the RHIC$\rightarrow$LHC \Jpsi~\textbf{un}suppression may yet turn out to be the promised smoking gun: Deconfinement implies colour conductivity, i.e. coloured partons can roam freely over distances much larger than the deconfinement scale. Such long distance travel is exactly what is needed in the regeneration model to explain how charm quarks produced in independent initial hard parton collisions can find each other at later times to recombine into a \Jpsi~during hadronisation. Note that coalescence of light flavour quarks (u/d/s) does not necessarily require long distance parton transport as light quarks are both abundant and can be generated locally at hadronisation.
However, given the charming history, some extra care and additional checks may be in order before declaring victory: to be quantitative, regeneration calculations need as input the total charm cross section (missing so far); a second unambiguous example for heavy quark recombination would be very helpful (eg $\rm B_c)$; the spectre of final state recombination at the hadron (D + D $\rightarrow$ \Jpsi~+ X) rather than parton level has to be very convincingly excluded (no easy feat given the many uncertainties in general associated with rate calculations in hadronic afterburners).

The original deconfinement signal, sequential quarkonia suppression (\RAA \Ups(1) $>$ \Ups(2) $>$ \Ups(3)), has been observed at LHC for the \Ups~family. Together with RHIC data (and smaller error bars), we may be able to see experimentally the effect of raising the temperature, if the melting pattern of different \Ups~states would change between the two energies. However, so far it seems that the \RAA~of \Ups~is very much comparable at RHIC and LHC (D\'{e}j\`{a} vu ?). Although the theory description of heavy quark suppression has advanced significantly, a complete and fully coherent description of all quarkonium data (\Jpsi~and \Ups~families) at SPS/RHIC/LHC needs significant further work (e.g. to include all of lQCD spectral functions, CNM effects like PDFs, energy loss, co-movers, hydro evolution, heavy quark diffusion, feed-down, ..).

\subsection{Collective Flow and Hydrodynamics}
Hydrodynamics as a tool to describe the space-time evolution of the expanding matter and extract its properties has been a mainstay of heavy ion physics essentially since its inception. It experienced a renaissance around 2010 when the importance of the initial state geometry and in particular its Event-by-Event fluctuations finally became clear. Since then, sophistication and quality of both the data (EbE flow measurements, event-shape engineering ESE, multi-particle correlations in phase and/or amplitude and/or rapidity,..) and theory (3+1 hydro, anisotropic hydro, bulk viscosity, freeze-out correction, hadron resonance gas (HRG) afterburners, magneto hydro dynamics,..) has been increasing to a remarkable level. The agreement between experiment and theory, often at the few \%~level even for the most arcane higher order multi-particle correlations, is nothing short of amazing. This is a major achievement in a field characterised originally as qualitative, at best. The increasing accuracy, together with novel observables and analysis methods like Bayesian multi-parameter fits, have led to a significant improvement in both constraining the sQGP (e.g. shear and bulk viscosity), as well as the detailed spatial structure of the initial state. The latter in particular has morphed from an input required for hydro into a fascinating physics topic of its own right, where the wave function of nucleons and the nucleus can be compared with ab initio calculations in the CGC framework and can be constrained from the data far beyond trivial geometry.

Amongst the many open questions and active areas of current research are: the elliptic flow of photons and heavy flavours, both measured to be typically larger than calculated by hydro; collectivity in small systems (discussed later); the hydro decoupling regime at intermediate \PT~of between a few and 10 GeV, where hadrons are increasingly  less well described by hydro  but not yet fully by pQCD and jets (note that the higher the mass, the larger the \PT~range over which hydro describes the data, which explains much of the "baryon anomaly").  Quark scaling (NCQ) remains an intriguing regularity in the data, but unlike hydro has unfortunately not progressed much from the initial back-of-the envelope scaling relations. Furthermore, NCQ scaling is still sometimes implored at low \PT~($\le 0.7 - 1$ GeV/quark) via the ad hoc use of transverse kinetic energy scaling, despite a consensus that standard hydro (+ CF freeze-out), which only knows about masses and momenta and nothing about quark content, is very successful and totally adequate in this momentum region. And finally, one open problem remains unsolved since the first realisation that $v_2$ and $v_3$ are about the same in very central collisions ($< 1 \% $ centrality \cite{ALICE:2011ab, ATLAS:2012at, CMS:2013bza}): Despite several attempts \cite{Luzum:2012wu,Denicol:2014ywa,Shen:2015qta,Plumari:2015cfa}, no model or parameter set has been able to reconcile these results with the expectation, i.e. $\epsilon_2 \approx \epsilon_3$ when fluctuations dominate, and therefore $v_3 < v_2$ with viscous damping $\eta/S>0$. As both data and the predictions seem very solid, the solution may be relevant even if it is more likely to be a thorny problem rather than a crack in the heavy ion standard model.
 
\subsection{Chiral Magnetic Signals}
QCD chiral symmetry, and its violation by quantum effects ("chiral anomaly"), leads in the presence of strong external magnetic fields to a number of charge specific multi-particle correlations commonly referred to as "Chiral Magnetic". Specific observables had been suggested and then unambiguously observed first at RHIC and later at LHC. Unfortunately, not all of the observed systematics (\PT, $\sqrt{s}$, correlation with the EP, ..) follow precisely the prediction, and also making the strongly Lorentz contracted magnetic field decay slowly enough to have any measureable effect at all is nontrivial.
Therefore, the leading background explanation of local charge conservation modulated by elliptic flow remains a very strong contender, even if it has its own problems with some of the observed regularities. A number of tests have been proposed which usually involve changing only one of the strength determining parameters at a time ($v_2$, magnitude or orientation of the magnetic field B). The first attempt (central U+U, $|B|=0, v_2>0$) gave ambiguous results; pPb ($<$$\vec{B}\mathrm{x}\vec{PP}$$> =0, v_2>0$) and ESE ($|B|=const, v_2\ne const$) seem to favour the background interpretation; eventually a future test at RHIC with isobar beams ($|B|\ne const, v_2=const$) could set a decisive limit if the question has not been settled by then. Today, while somewhat on the defensive, the various chiral magnetic signals remain amongst the most enigmatic and potentially most important results from heavy ion collisions; with significant experimental signals to work with and a solid and fundamental relation to  QCD, the issue merits more effort towards a definitive answer.

\subsection{Energy Loss and Jet-quenching}
Since the initial RHIC discovery of inclusive single particle high \PT~suppression, the study of energy loss has made enormous quantitative and qualitative progress on both the experimental and the theory fronts. The "stopping power of sQGP", $\hat{q}/T^3$, is now known to about 30\% and the predicted mass dependence has been investigated with heavy flavour mesons. Since actual reconstructed jets have entered the toolbox of jet-quenching, the high \PT~suppression has been unambiguously connected to energy loss (from the jet \PT~imbalance), and increasingly detailed studies of the medium modified jet fragmentation and splitting functions has given insight into the dynamics of the jet medium interaction and the approach to jet "thermalisation". 
Amongst the open questions is identifying the response, if any, of the medium to the concentrated local energy deposition (Mach Cone, collective wake,..). Future high precision measurements with more statistics, in particular in the golden $\gamma$-jet channel, and new detectors (sPHENIX), are needed to confirm the ever so slight hint of a non-trivial temperature dependence of the stopping power (RHIC $\rightarrow$ LHC) and nail its parton type, parton energy, and path-length dependence.
And finally, somewhat unsettling remains, at least to this author, the miraculous cancellations required concurrently in spectral slopes, fragmentation functions, and energy loss coefficients, in order to explain the nearly invariant magnitude and shape of \RAA~for different parton flavours ($g\approx u/d/s\approx c\approx b\approx $~\Jpsi) and at different energies ($0.2\approx2.8\approx5$ TeV).

\subsection{Small is Beautiful}
The first LHC discovery, and arguably still the most unexpected one, was announced in 2010 when CMS presented evidence of a long range, near side ridge in particle correlations observed in high multiplicity pp collisions. The signal was so unexpected that the CMS spokesperson at the time announced it with the warning ".. we didn't succeed to kill it  .. [and] therefore expose our findings to the scrutiny of the scientific community.." \cite{cmsridge}. The discovery spawned a menagerie of theoretical explanations \cite{ Li:2012hc}, spanning the gamut from a priori very unlikely to outright weird. Two serious contenders remain today, i) initial state quantum correlations as calculated by CGC and ii) final state interactions leading to collective flow described with hydrodynamics.

Largely set aside as an unexplained curiosity after two years, the ridge came back with a vengeance in pPb collisions. It could no longer be ignored, and the full machinery only recently developed for the study of collectivity in nuclear collisions was brought to bear on the phenomenon. As of today, all results clearly and unambiguously point into the same direction: The correlations seen in small (pp/pA/dA/HeA) systems are truly collective (i.e. $v\{4\} \approx v\{6\}\approx v\{8\}$..) and bear all the hallmarks of hydrodynamic flow as previously seen only in nuclear collisions (higher harmonic azimuthal flow, mass and \PT~differential flow, factorisation violation, harmonic correlations, multiplicity dependence,..).  A crucial test, and a triumph for the hydro explanation, was the correlation measurement at RHIC using d and $^3$He projectiles to introduce particular initial state geometries. The $2^{nd}$ and $3^{rd}$ harmonic flow components were exactly as predicted by hydro codes, whereas the CGC model is still toiling to find a convincing explanation.

In parallel, other results which in the past had been associated exclusively with nuclear collision and hot/dense matter formation have been looked for, and found, in pp and pA reactions:
 from strangeness enhancement and increasingly thermal (grand canonical) particle production to HBT radii, charge balance functions, and inclusive \PT~spectra which scale with multiplicity, particle mass,  and/or momentum exactly as expected for a system undergoing collective expansion and cooling \cite{Song:2017wtw, Dusling:2015gta, Loizides:2016tew}.
A notable exception remains jet-quenching (high \PT~suppression), which, within current experimental uncertainties of order 10-20\%, is nowhere to be found in pPb. All other phenomena, including the elliptic flow, develop smoothly and without any apparent threshold, starting with multiplicities at or below MinBias pp. As far as one can tell, two particles is company, three is a (collective) crowd.

Were it not for the difficulty to conceive of hydrodynamic behaviour not only in small and dense, but also in small and dilute systems, where hydro is thought to be well outside its range of applicability (mean free path/system size $<<1$ and \#collisions/particle $<<1$, see however \cite{Romatschke:2017vte}), the well-motivated but increasingly Ptolemaic CGC explanation would have had to concede defeat a while ago, given the clear preponderance of evidence.
While hydrodynamic codes can and are being used in ever smaller systems, down to pp at LHC, and actually give very consistent and coherent results well in line with the data \cite{Weller:2017tsr}, the debate does (and should) continue until we do understand the causes and the actual dynamics behind what is yet another case for an unreasonably successful model (hydro in small dilute systems).

Insight into the dynamics should come from checking how well microscopic transport models describe both large/dense and small/dilute systems and the transition between them.
One of the widely used transport codes, AMPT, can reproduce surprisingly well most two- and multi-particle correlation measures for all pA and AA systems at both RHIC and LHC (it is less good in describing \PT~spectra and particle ratios). Besides using a reasonable, if not unique, initial state geometry, which crucially however included EbE fluctuation long before they were generally known to be relevant \cite{Chen:2004dv}, AMPT contains a dense initial parton state with a fairly simplistic but very effective parton scattering phase, followed by coalescence hadronisation and HRG afterburner. Its amazing success to describe, even predict, collective correlation signals in AA despite its obvious simplicity and theoretical shortcomings was attributed to a kind of duality, as hydro emerges as the statistical long wavelength limit of a strongly interacting microscopic theory, largely independent of the particular implementation of the microscopic dynamics. The perception of AMPT as a respectable transport substitute for hydro - wrong in theory but right in practice - changed when it was realised that the AMPT way of generating azimuthal "flow" is very far from a pressure driven hydro limit, at least in dilute systems like pA or even peripheral AA \cite{He:2015hfa}. Dubbed "Escape Mechanism", but better qualified as "density tomography", the AMPT transport generates an image of the initial state density distribution with single parton scatterings, which obviously leads to a depletion of particles in the direction of increased matter density. Hydro requires several collisions/particle to transform density inhomogeneity into pressure, whose gradients in turn drive the azimuthal modulation of particle distributions. On the contrary, the direct "X-ray" mechanism of AMPT works perfectly well even in very dilute systems where the majority of particles "escape" without interaction, as the remaining few which do scatter still take a faithful image of the initial state matter density, albeit one with potentially very low contrast. Note that even for very dilute systems (\#collisions/particle $<<1$), density tomography as implemented by AMPT remains fully collective (the density imprint is on the single particle angular distribution, which is a necessary and sufficient condition for collectivity) and seems to reproduce fairly well not only various flow harmonics but also more complex hydro signals like plane and amplitude correlations or non-linear mode mixing. For the time being, it is unclear (to this author) if pressure tomography as implied by hydro and density tomography as implied by AMPT are two sides of the same coin (the high and low density limits, respectively), or are two different physics mechanism with different observable consequences (say in directed photon flow or non-liner mode mixing). The former would make the question moot, the latter would open it to experimental examination.

The fact that apparently all hadronic reactions, if measured with appropriate sensitivity, show collective features which (so far) are qualitative indistinguishable from hydro flow, and quantitativly  scale with dN/dy as expected (two particle correlation amplitude $\propto$ N$^2$ or $v_n \approx$ constant), is only hesitantly accepted as relevant and incorporated in what used to be "heavy ions only" physics.  On one hand, we should not be too surprised if macroscopic hydro morphs smoothly and without discernible threshold into a microscopic transport theory (referred to as low-density limit); on the other hand, it was certainly not to be expected that the validity of second order viscous hydrodynamics extends all the way down to systems containing a few particles only. 

In any case, the old paradigm that we study hot and dense matter properties in heavy ion collisions, cold nuclear matter modifications in p-nucleus, and use pp primarily as comparison  data, appears no longer sensible. We should examine a new paradigm, where the physics underlying soft 'collective' signals (including apparently hadronisation!) is the same in all high energy reactions, from \ee~to central AA. This physics may be a generic property of all strongly interacting many-body systems ($N \geq 2 ?$), even if it is obvious and dominant only in AA and barely discernible in pp. The same line of thought has taken root also with some of our colleagues on the HEP side, as shown in a recent preprint entitled "Thermodynamical String Fragmentation" \cite{ Fischer:2016zzs} where an attempt is made to incorporate thermal model hadronisation into the Phythia pp event generator. As the author remarks: "The understanding of soft hadronic physics is changing under the onslaught of LHC. .. we have an interesting and challenging time ahead of us, where some of the most unexpected new LHC observations may well come in the low-\PT~region rather than the in high-\PT~one."

By looking  at  small systems, once more we have found, as stated 12 years ago at a different occasion, that the \emph{".. hot, dense matter ..[is] .. quite different and even more remarkable than had been predicted .."}. 

\subsection{The Low Energy Frontier}
In the dash to new machines and higher energy, the question of the onset - exactly when (in time) and where (in energy or volume) we first produced the QGP - has been left behind, unanswered. We know today about a number of intriguing structures ("steps, kinks, dales", local mini- or maxima) in specific observables when plotted as a function of energy and/or system size. However, the hints appear at different places (some around $\sqrt{s}~7-8$ GeV, others at 10-20 GeV) and most observables emerge or change very smoothly. For the time being, there are many hints, but no conclusive picture. It may yet turn out that there is actually no clearly identifiable onset, that the transition is continuous and gradual in terms of volume (starting with pp) or energy density, and possibly even happens at different places for different observables.  This would after all mirror the characteristics of the Equation-of-State, which we know from lQCD (at small $\mu_B$) to be a smooth cross-over rather than an actual change of phase. Besides the onset, the other landmark still missing in the phase diagram  is the conjectured critical point, tentatively located at large baryon density and therefore at low beam energy. 

Both questions, onset and critical point, are addressed with the past and future beam energy scans at RHIC and an energy-volume mapping ongoing at the SPS. Sometime early in the next decade the new machines SIS100 at GSI/FAIR, NICA at JINR, and an upgrade proposed for  J-PARC, will enter the race at the low energy frontier, to search for the onset, the critical point, and more general study matter properties at high baryon density. The future competition at this end of the phase diagram seems fierce and plenty, and with the low hanging fruits presumably picked already at the dawn of the heavy ion program, both a very performant machine and a first-class experimental program is required for success.  

\section{Summary}
The field of ultra-relativistic heavy ion physics has seen incredible rapid progress in a mere three decades: covering 4 orders of magnitude in total kinetic energy from OAu at the AGS to PbPb at the LHC, from reusing borrowed HEP equipment to building state-of-the-art dedicated heavy ion detectors, from a few dozen pioneers to several thousand practitioners, from the periphery to the center of contemporary nuclear physics. Along the way, we have explored the previously uncharted territory of hot and dense nuclear matter: We have measured with increasing precision its macroscopic properties and many transport coefficients (EOS, $c_S, \eta/S, \xi/S$, D, $\hat{q}, \hat{e}$); we have good evidence for deconfinement from \Jpsi~regeneration (colour conductivity) and sequential \Ups~suppression (resonance melting); we have some indirect evidence for chiral symmetry restoration ($\rho$ melting, strangeness enhancement). And we also learned that the d.o.f. of the Quark-Gluon Plasma are definitely \textbf{not} (quasi-free) quarks and gluons.

The forthcoming high luminosity runs and detector upgrades at both RHIC and LHC (and possibly the SPS ?) should significantly increase data quality and data quantity to advance on the  precision frontier on a variety of existing (e.g. transport coefficients) or new (e.g. low mass lepton pairs at LHC)  signals; hopefully to the level required to see, or to meaningfully limit, for example a change in the quarkonia suppression pattern from RHIC to LHC  or a nontrivial temperature dependence of jet-quenching. 
On a deeper level, much also remains to be discovered: Primarily, \emph{what are} actually the relevant degrees of freedom?  What is the underlying dynamics, which makes thermalisation (if really achieved?) seemingly so fast, including in small and possibly even in dilute systems? What mechanisms generate the statistical particle ratios, from \ee~to central AA, or produce (the mirage of?) a strongly interacting liquid in tiny systems containing a mere handful of particles? Is there a critical point and a first order phase transition at high baryon density? Is there an onset of the sQGP, and how does collectivity and macroscopic behaviour emerge with system size and/or energy density? 

Every time in the past when we thought our research has finally matured, passing from youthful but unsteady days of rapid discoveries to a more staid pace of increasing precision and understanding, some major surprise has given the field a new direction, new questions, and a new impetus. 
When dealing with the strong interaction in a new regime where it \emph{is} strong, and therefore to some extent unpredictable, new insights and even surprises, presumably, should be expected. We can therefore look forward with anticipation to the new results presented at this 2017 edition of the Quark Matter Conference, and at the next QM, and the next after that,...


\bibliographystyle{elsarticle-num}
\bibliography{JSQM17}







\end{document}